\documentclass[a4paper,12pt]{article}
\usepackage{indentfirst} 
\usepackage{url}         
\usepackage{upquote}     
\usepackage{xcolor}      
\usepackage{orcidlink}   
\usepackage{fontawesome5}
\usepackage{amsmath}
\usepackage{multirow}

\usepackage{etoolbox}                       
\AtBeginDocument{\hypersetup{hidelinks}}    

\usepackage{authblk}                        

\usepackage{geometry}			
\usepackage{amsmath,amsfonts,amssymb,amsthm,mathtools}

\usepackage{graphicx}					
\graphicspath{{figures/}}		        
\usepackage{caption}                    
\usepackage{array,tabularx,tabulary,booktabs}	

\usepackage{cite}						

\usepackage{nomencl}                    

\begin{document}

\title{Analysis and Control of Acoustic Emissions \\ from Marine Energy Converters}
\author[1]{Jiaqin He}
\author[1]{Max Malyi\orcidlink{0000-0002-1503-9798}\thanks{Corresponding author: \url{Max.Malyi@ed.ac.uk}}}
\author[1]{Jonathan Shek\orcidlink{0000-0001-5734-2907}}

\affil[1]{Institute for Energy Systems, School of Engineering, The University of Edinburgh, Edinburgh, UK}

\date{August, 2025}

\maketitle

\begin{abstract}
Environmental licensing related to underwater acoustic emissions represents a critical bottleneck for the commercial deployment of marine renewable energy. This study presents a control engineering framework to mitigate acoustic risks from tidal current converters without compromising project viability. A MATLAB/Simulink model of a tidal current converter was utilised to evaluate two distinct mitigation tiers: (1) architectural modification, comparing a geared induction generator against a direct-drive permanent magnet synchronous generator, and (2) operational control, analysing the impact of switching frequencies and maximum power point tracking coefficient ($K_{opt}$) tuning. Results indicate that lowering switching frequencies ($F_s$) is ineffective, increasing power electronic losses by over 2000\% with negligible acoustic benefit. Conversely, the direct-drive permanent magnet synchronous generator architecture reduced sound pressure levels by approximately 10 dB re 1 $\mu$Pa, effectively eliminating mechanical tonal noise. For existing geared systems, de-tuning the $K_{opt}$ coefficient by a factor of 1.2 reduced the probability of exceeding temporary threshold shift limits for marine mammals, with a quantified energy yield reduction of 3.58\%. These findings propose a hierarchical mitigation strategy: selecting direct-drive topologies for acoustically sensitive sites, and utilising maximum power point tracking coefficient based power curtailment as a transient operational mode during critical biological migration periods.
\end{abstract}

\begin{figure}[h!]
    \centering
    \includegraphics[width=0.8\textwidth]{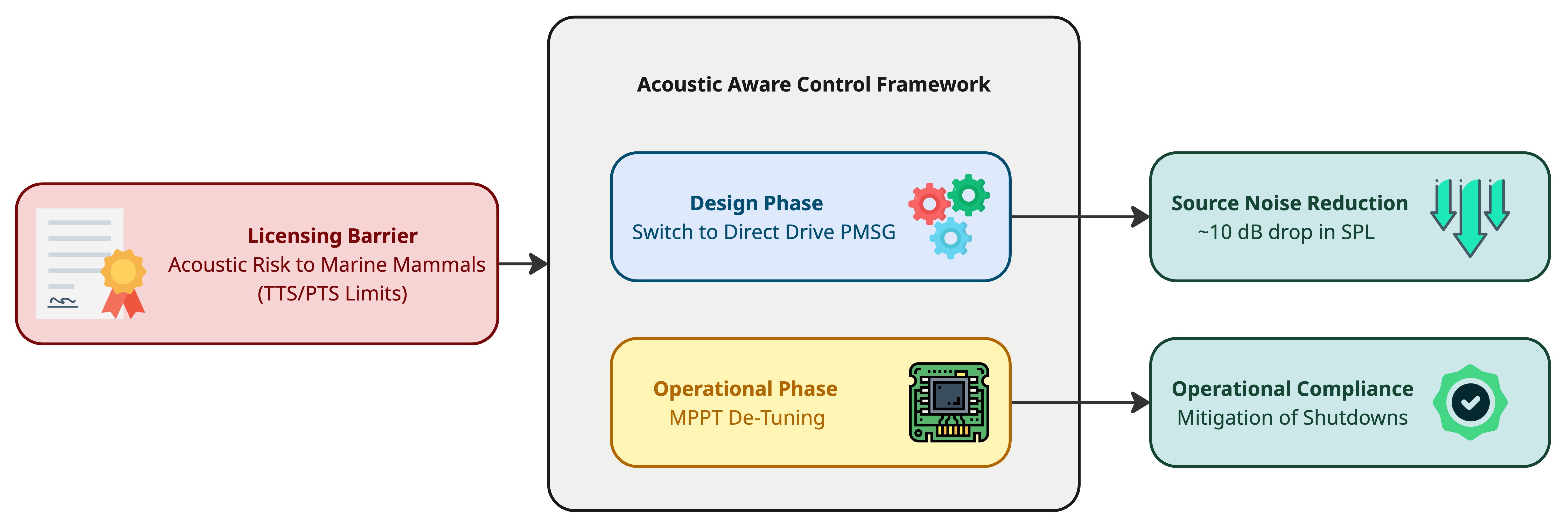}
\end{figure}

\vspace{0.1cm}
\noindent\rule{\linewidth}{0.3pt}
\vspace{0.1cm}

\noindent\small\textbf{Keywords:} marine energy, acoustic emissions, direct-drive PMSG, control optimisation.


\newpage

\section{Introduction}
The global imperative to mitigate climate change has accelerated the transition towards sustainable energy systems, with a projected 50\% increase in global energy consumption between 2012 and 2040 necessitating robust renewable alternatives~\cite{conti2016international}. Within this landscape, marine energy, encompassing tidal stream and wave energy, represents a vast, under-utilised resource. It is estimated that the kinetic and potential energy of the oceans could satisfy up to 20\% of the UK's electricity demand~\cite{SDC2007}. Unlike wind and solar, tidal stream energy offers high power density and astronomical predictability~\cite{vennell2013exceeding}, making it a critical candidate for providing base-load stability to decarbonised grids~\cite{clarke2010survey}.

However, despite this potential, the commercial deployment of marine energy converters lags significantly behind other renewable sectors~\cite{lepper_acoustic_2012}. The industry faces a complex nexus of challenges, including high levelised costs, lack of design convergence, and hostile operating environments. Yet, one of the most significant non-technical barriers to deployment is environmental consenting and licensing~\cite{magagna2015ocean}. Regulators strictly enforce compliance with environmental standards, particularly regarding the impact of underwater acoustic emissions on marine life~\cite{jmse8110879}.

\subsection{The Engineering Challenge of Acoustic Compliance}
The deployment of tidal current converters (TCCs) inevitably introduces anthropogenic noise into the marine environment. This noise is not merely a by-product but a critical operational constraint. Marine mammals rely heavily on sound for communication, navigation, and foraging. Excessive acoustic emissions can lead to severe biological consequences, ranging from behavioural disturbance and auditory masking to physiological damage such as temporary threshold shifts (TTS) and permanent threshold shifts (PTS)~\cite{peng2015noise,tougaard2015underwater}.

From an engineering perspective, acoustic compliance effectively functions as a "Go/No-Go" gate for site licensing. In narrow channels where tidal energy extraction is most efficient, turbine arrays risk creating acoustic barriers that deter marine species from critical habitats or migration routes~\cite{wilson_use_2013}. Consequently, the inability to guarantee acoustic safety can lead to permit denials or severe operational curtailment.

Current mitigation strategies are often passive, relying on site avoidance or the use of acoustic deterrent devices to repel animals from the turbines. These methods are suboptimal; they restrict the available resource potential and introduce additional noise pollution. There is a critical need for active, source-based mitigation strategies. By treating acoustic emission as a controllable system state, analogous to voltage or torque, engineers can design TCCs that inherently comply with environmental thresholds without requiring total system shutdown.

\subsection{Acoustic Source Characterisation and Control Variables}
\begin{figure}[!htbp]
    \centering
    \includegraphics[width=0.5\linewidth]{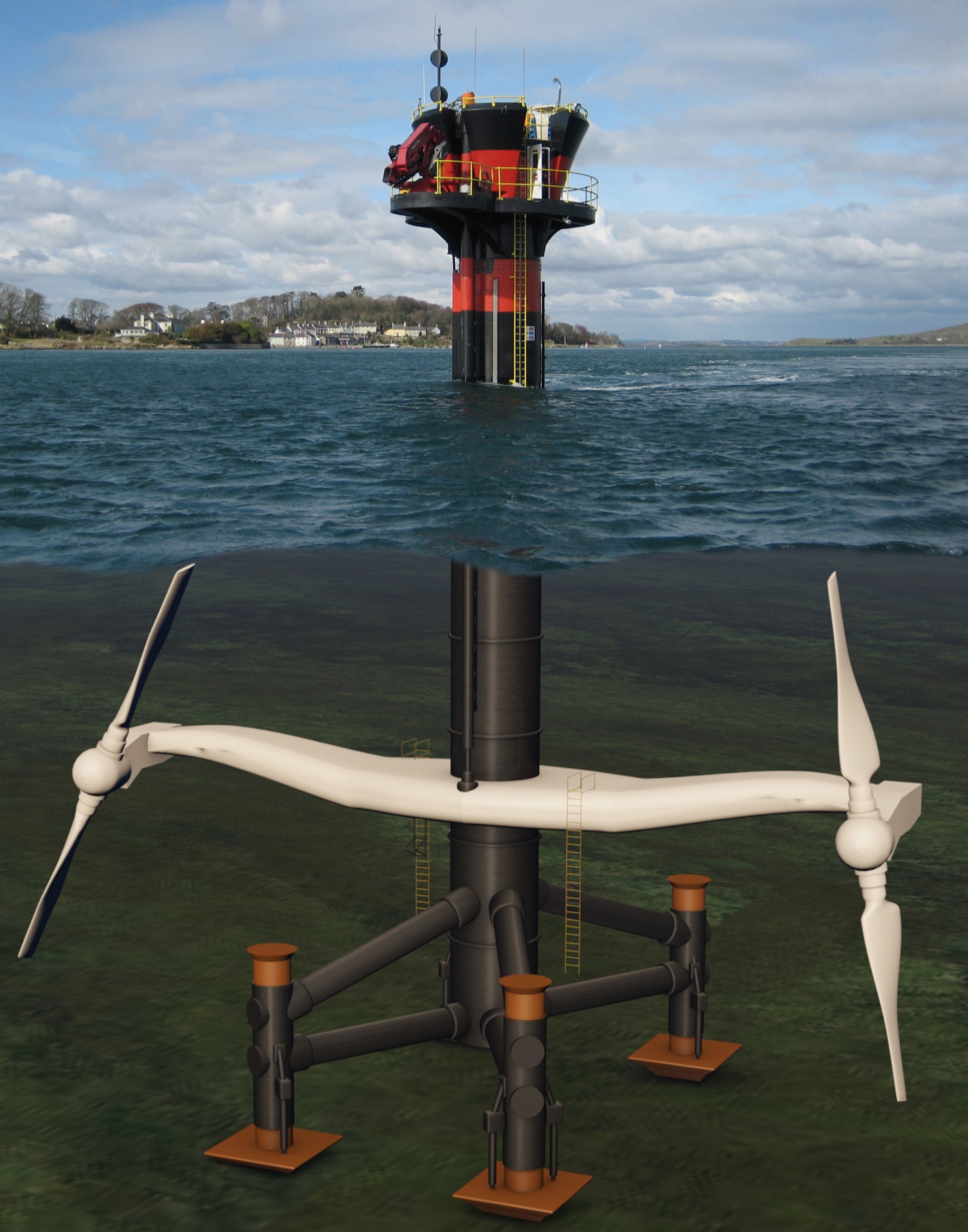}
    \caption{The SeaGen tidal turbine, a commercial-scale horizontal-axis TCC deployed in Strangford Lough, Northern Ireland~\cite{turbines2011seagen}.}
    \label{fig:seagen}
\end{figure}

To control acoustic emissions, one must first characterise the generation mechanism. The acoustic signature of a horizontal-axis TCC (e.g., Fig.~\ref{fig:seagen}) is a complex aggregate of hydrodynamic and mechanical sources~\cite{lloyd2011noise}:

\begin{itemize}
    \item \textbf{Inflow Turbulence (Hydrodynamic):} Interaction between the turbulent water flow and the rotating blades generates broadband noise, typically dominant at lower frequencies~\cite{wang2007experimental}. This source is intrinsically linked to the rotor's angular velocity ($\omega$) and the Tip Speed Ratio ($\lambda$).
    \item \textbf{Mechanical Noise:} Components within the drivetrain, specifically the gearbox and generator, emit tonal noise at specific frequencies related to gear meshing and electromagnetic switching. These tones often overlap with the most sensitive hearing ranges of marine mammals (e.g., porpoises and seals)~\cite{richards2007underwater}.
\end{itemize}

The magnitude of these emissions is governed by the turbine's operational point. The kinetic power $P$ extracted by the turbine is defined by:

\begin{equation}
\label{eq:kinetic_power_fluid}
P = \frac{1}{2} \rho A U^{3} C_p(\lambda, \beta),
\end{equation}

where $\rho$ is fluid density, $A$ is rotor swept area, $U$ is flow velocity, and $C_p$ is the power coefficient, which is a non-linear function of the tip speed ratio $\lambda$ and blade pitch angle $\beta$~\cite{bahaj2003fundamentals}.

Conventional control strategies, such as maximum power point tracking (MPPT), aim solely to maximise $C_p$. However, this study posits that the control objectives must be expanded. By manipulating control variables, specifically the generator topology, switching frequencies of power electronics, and the MPPT gain coefficients, it is possible to alter the acoustic profile of the turbine. For instance, reducing rotational speed ($\omega$) to operate at a sub-optimal $\lambda$ (de-rating) reduces hydrodynamic noise, albeit at the cost of energy yield. Quantifying this trade-off is essential for establishing the cost of environmental compliance.

\subsection{Research Objectives}
This study aims to bridge the gap between marine ecology and control engineering by developing a framework for acoustic-aware control. Building upon a validated MATLAB/Simulink model of a grid-connected tidal current conversion system~\cite{sousounis2016modelling}, we simulate the acoustic emissions of a TCC under various operational strategies.

Specific objectives include:
\begin{itemize}
    \item \textbf{Source Decomposition:} To quantify the individual acoustic contributions of inflow turbulence, the gearbox, and the generator, and evaluate their respective impacts on marine mammal auditory thresholds (TTS/PTS).
    \item \textbf{Architectural Evaluation:} To compare the acoustic performance of a conventional geared induction generator against a direct-drive permanent magnet synchronous generator (PMSG), assessing the noise reduction potential of eliminating the gearbox.
    \item \textbf{Control Optimisation:} To investigate the efficacy of operational interventions, specifically varying the switching frequency ($F_s$) of the power converters and de-tuning the MPPT coefficient ($K_{opt}$), in mitigating acoustic risk.
    \item \textbf{Cost-Benefit Analysis:} To quantify the energy yield penalty associated with these mitigation strategies, providing developers with data-driven trade-offs between power production and environmental compliance.
\end{itemize}

By shifting the focus from impact assessment to active control, this research demonstrates how engineering innovations can resolve regulatory bottlenecks, thereby facilitating the wider deployment of clean marine energy technologies.

\section{Methodology}

\subsection{Operational Constraints: Acoustic Thresholds}
To establish the operational boundaries for the control system, this study utilises biological damage criteria for marine species prevalent in UK waters, specifically marine mammals such as harbour porpoises and seals, among others~\cite{macleod2007habitat}. Audiograms and generic hearing thresholds (GTV) were compiled from literature~\cite{chapman1973field,awbrey_low_1988,kastelein_audiogram_2003,kastelein_underwater_2009,kastelein2010effect,szymanski_killer_1999,nachtigall_shipboard_2008}.

\begin{figure}[!htbp]
    \centering
    \includegraphics[width=0.6\linewidth]{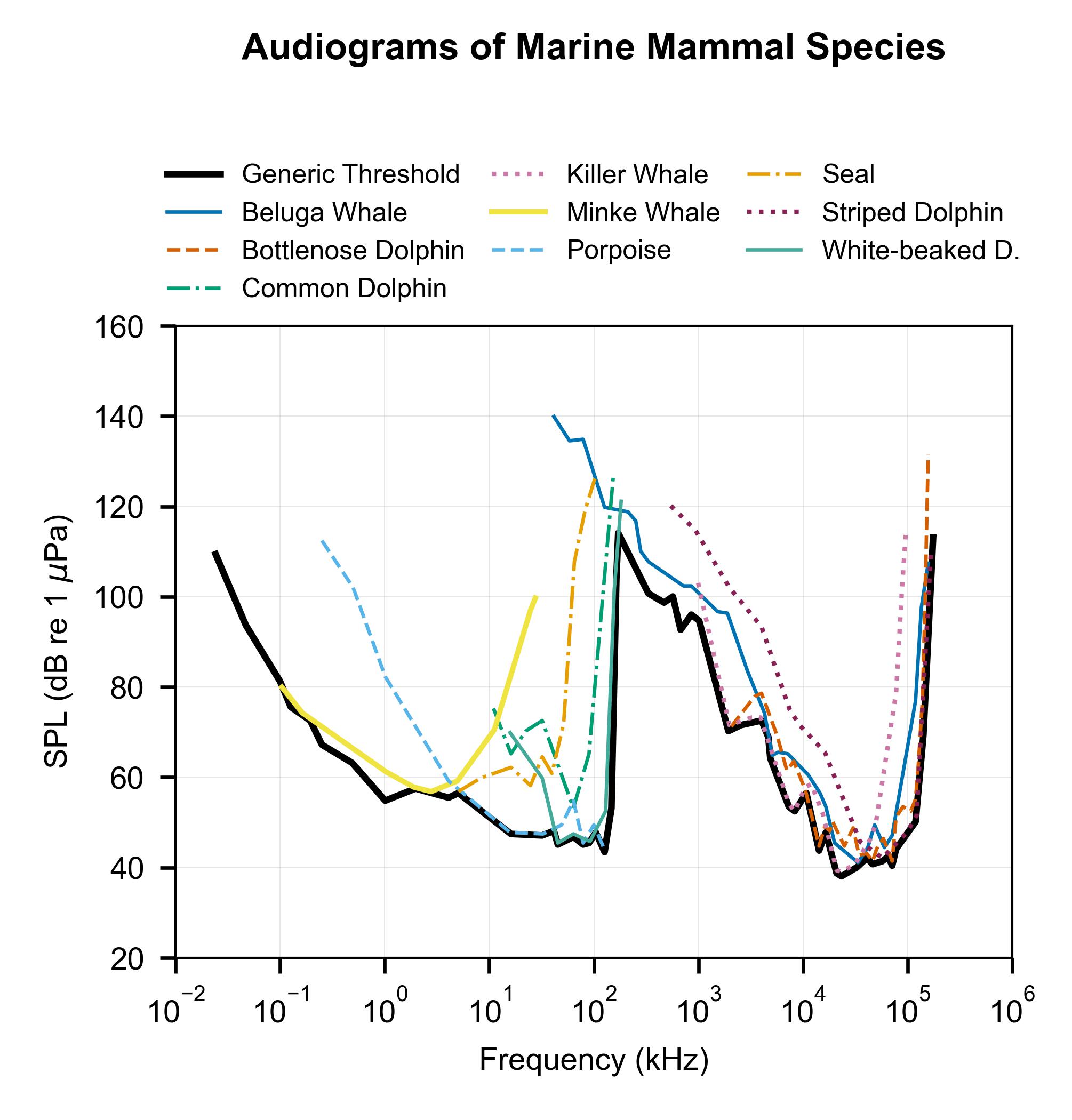}
    \caption{Audiograms of target marine mammal species and the derived generic hearing threshold (GTV) used as the baseline for acoustic impact assessment.}
    \label{fig:audiograms}
\end{figure}

The engineering constraints are defined by the onset of TTS and PTS. Following the exposure models by Heathershaw et al.~\cite{heathershaw2001environmental} and Richards et al.~\cite{richards2007underwater}, the limit thresholds are calculated as:

\begin{equation}
\label{eq:TTS_threshold}
TTS = GTV + 75 - 10 \log_{10} \left( \frac{T}{28800} \right),
\end{equation}

\begin{equation}
\label{eq:PTS_threshold}
PTS = GTV + 95 - 10 \log_{10} \left( \frac{T}{28800} \right),
\end{equation}

\noindent where $T$ is the daily exposure duration in seconds. For a grid-connected tidal turbine operating continuously in base-load conditions, a worst-case exposure scenario ($T \ge 28,800$ s/day) is assumed. Consequently, the logarithmic duration term approaches zero, setting the static engineering limits at $GTV + 75$ dB (TTS) and $GTV + 95$ dB (PTS).

\subsection{Modelling of Hydrodynamic Noise (Inflow Turbulence)}
Hydrodynamic noise, resulting from the interaction between unsteady inflow turbulence and the rotating blades, acts as the dominant low-frequency noise source. This was modelled using the empirical spectral method adapted by Lloyd et al.~\cite{lloyd2011modelling} from the original formulation by Blake~\cite{blake1984aero}.

The mean square pressure $\overline{p^{2}}$ at a distance $r$ is derived from the pressure spectrum $\phi_{p}(r, \omega)$:

\begin{equation}
\label{eq:mean_square_pressure}
\overline{p^{2}} = \frac{1}{2} \phi_{p}(r, \omega) \omega,
\end{equation}

\noindent where $\omega$ is the angular frequency. The pressure spectrum is related to the turbulence spectrum $\phi_{t}(\omega)$ by:

\begin{equation}
\label{eq:pressure_spectrum}
\phi_{p}(r,\omega) = \frac{k_{0}^{2} \cos^{2}\beta}{16\pi^{2}r^{2}} \phi_{t}(\omega),
\end{equation}

\noindent where $k_{0} = \omega/c$ is the acoustic wavenumber ($c$ is the speed of sound in water). The turbulence spectrum $\phi_{t}(\omega)$ incorporates the rotor geometry and the Sears function to account for the aero-acoustic response of the blade sections:

\begin{equation}
\label{eq:turbulence_spectrum}
\phi_{t}(\omega) = \frac{2}{3}\pi^{2}\Lambda_{R}\rho^{2}U_{T}^{2}C_{T}^{2}\frac{D}{2} |A_{s}|^2 \overline{u^{2}} F_{\Lambda} |S_{e}|^2,
\end{equation}

\noindent where $\Lambda_R$ is the radial length scale of turbulence, $\rho$ is fluid density, $U_T$ is the tip velocity, $C_T$ is the tip chord length, and $D$ is the rotor diameter. The term $|S_e|^2$ represents the approximated Sears function, which models the lift response to sinusoidal gusts:

\begin{equation}
\label{eq:sears_function}
|S_{e}(\kappa)|^2 \approx \frac{1}{1 + 2\pi \kappa}, \quad \text{where } \kappa = \frac{\omega C_T}{2 U_T}.
\end{equation}

This formulation ensures that the simulated acoustic profile accurately reflects the physical dependency of noise on rotor speed ($\Omega$) and tip speed ratio ($\lambda$).

\subsection{Modelling of Mechanical Drivetrain Noise}
Mechanical noise sources were modelled based on empirical industrial machinery standards established by Bruce et al.~\cite{bruce2007sound}. The sound power levels ($SL$) for the gearbox and generator were calculated as:

\begin{equation}
\label{eq:sl_gearbox}
\small
SL_{\text{gear}} = 86 + 3 \log_{10} (rpm_{s}) + 4 \log_{10}(P_{kW}) + 10 \log_{10}(S_{geom}),
\end{equation}

\begin{equation}
\label{eq:sl_generator}
SL_{\text{gen}} = 80 + 10 \log_{10}(P_{MW}) + 6.6 \log_{10}(rpm_{gen}),
\end{equation}

\noindent where $rpm_{s}$ is the shaft speed, $P$ is power (in kW or MW as denoted), and $S_{geom}$ represents the geometric surface area of the casing.

Since these empirical relationships are derived for airborne acoustics (reference pressure $20 \mu$Pa), the results were converted to underwater sound pressure levels ($SPL_{water}$, reference $1 \mu$Pa) by accounting for the difference in acoustic impedance between air and water (approximately $61.5$ dB correction) and the propagation loss equation~\cite{lloyd2011noise}. No nacelle damping factor was applied, representing a worst-case unshielded scenario.

\subsection{Control Strategies and System Architecture}
The TCC operation was simulated using a grid-connected tidal current conversion system model developed in \\ MATLAB/Simulink~\cite{sousounis2016modelling}.

\begin{figure}[!htbp]
    \centering
    \includegraphics[width=0.75\linewidth]{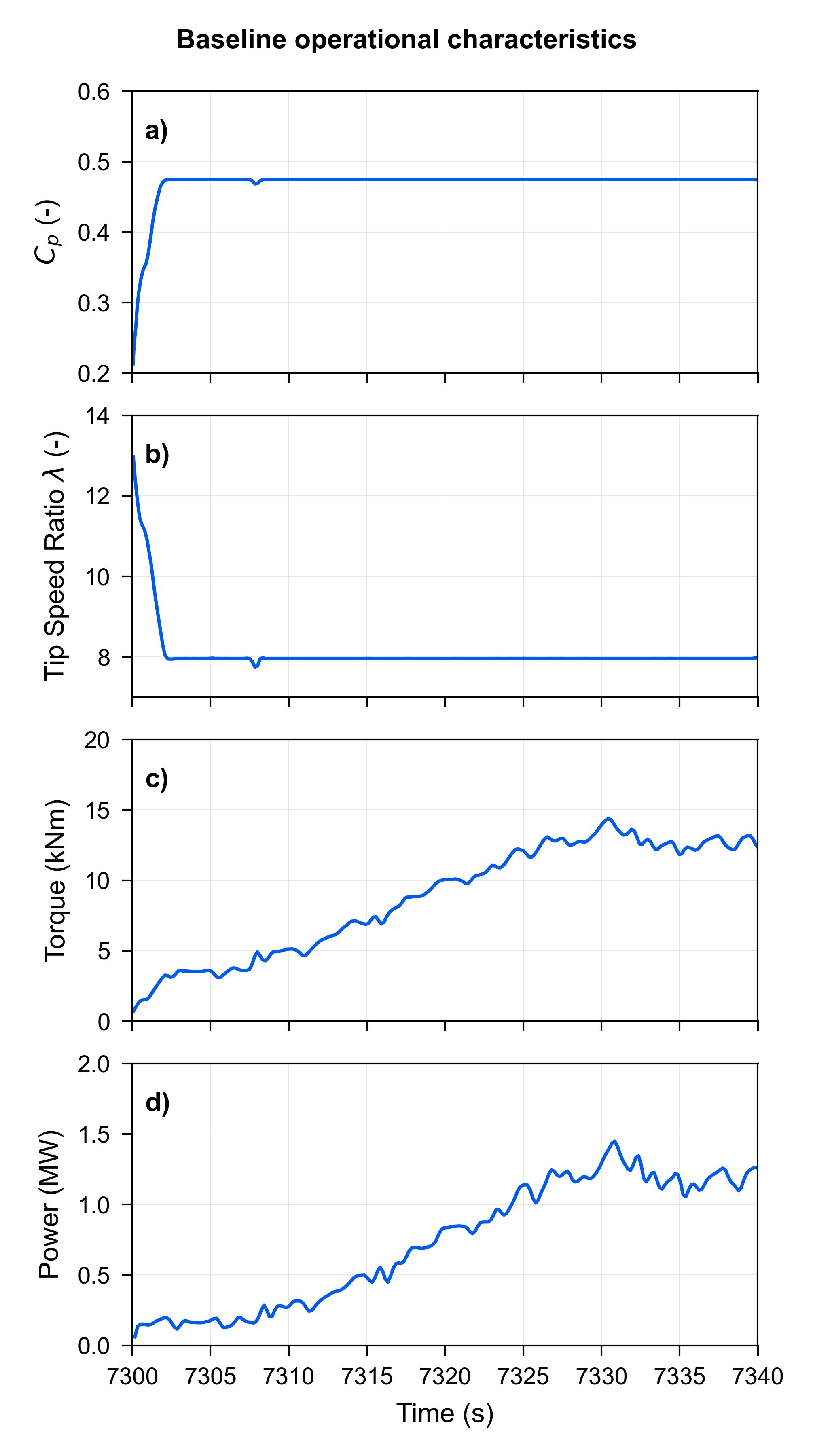}
    \caption{Baseline operational characteristics of the simulated TCCS: (a) power coefficient $C_p$, (b) tip speed ratio $\lambda$, (c) mechanical torque, and (d) power output. The simulation demonstrates the dynamic response of the control system to turbulent inflow.}
    \label{fig:timeseries}
\end{figure}

The baseline model (Fig.~\ref{fig:timeseries}) represents a 1 MW horizontal-axis turbine with a rotor diameter of 23m. To evaluate acoustic mitigation, three specific engineering interventions were simulated:

\subsubsection{Power Electronics Switching Frequency ($F_s$)}
The switching frequency of the generator-side converter was varied from 1 kHz to 3 kHz in 500 Hz increments. This strategy aimed to assess whether shifting the harmonic content of the electromagnetic noise could reduce the aggregate SPL without incurring excessive thermal losses in the filter components.

\subsubsection{MPPT Coefficient Tuning ($K_{opt}$)}
The supervisory controller regulates the generator speed to maintain optimal $C_p$ via the law $T_{em} = K_{opt} \cdot \omega^2$. The gain coefficient $K_{opt}$ was treated as a variable control parameter (varied by factors of 0.8 to 1.2). This effectively de-rates the turbine hydrodynamically, forcing it to operate at a sub-optimal tip speed ratio ($\lambda$) to reduce rotational noise at the expense of energy yield.

\subsubsection{Drivetrain Architecture: Geared vs. Direct-Drive}
To evaluate design-stage mitigation, the system topology was altered from a standard geared induction generator (reference) to a direct-drive PMSG.
\begin{itemize}
    \item \textbf{Reference Model:} 3-stage gearbox, 4-pole induction generator ($p=2$).
    \item \textbf{Direct-Drive Model:} Gearbox removed, PMSG with high pole count ($p=30$), rated torque increased to 955 kNm, and rated speed reduced to 15 rpm~\cite{haikonen2013characteristics,lee2012case}.
\end{itemize}
This structural change eliminates the gear meshing tones entirely, isolating the aerodynamic and electromagnetic noise components.

\section{Results}

\subsection{Baseline Acoustic Performance}
\begin{figure}[!htbp]
    \centering
    \includegraphics[width=0.75\linewidth]{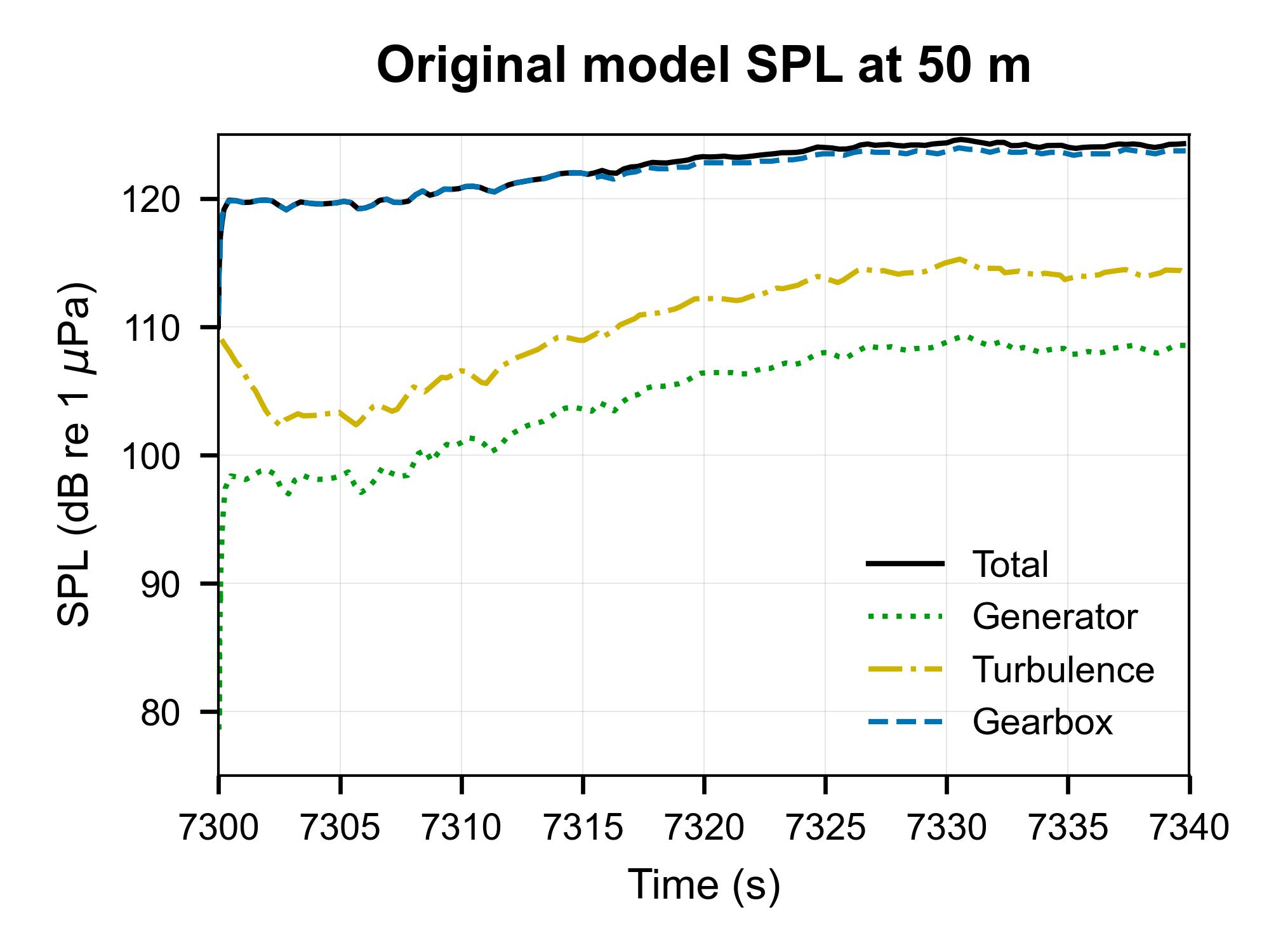}
    \caption{Time-series of SPL components for the reference geared TCC at 50 m distance.}
    \label{fig:spl-original-model-50m}
\end{figure}

The simulation of the reference TCC (geared induction generator) under standard control strategies established a baseline acoustic profile. Fig.~\ref{fig:spl-original-model-50m} illustrates the time-domain evolution of the SPL at a receiver distance of 50 m. The total SPL stabilises at approximately $124.2$ dB re 1 $\mu$Pa as the turbine reaches its rated operation. Decomposition of the acoustic signal reveals that mechanical noise from the gearbox is the predominant source, contributing consistently between $122$ and $124$ dB re 1 $\mu$Pa. In contrast, the generator noise remains significantly lower ($\sim 108$ dB re 1 $\mu$Pa), while inflow turbulence noise fluctuates with the tidal current velocity, averaging $114$ dB re 1 $\mu$Pa.

\begin{figure}[!htbp]
    \centering
    \includegraphics[width=0.75\linewidth]{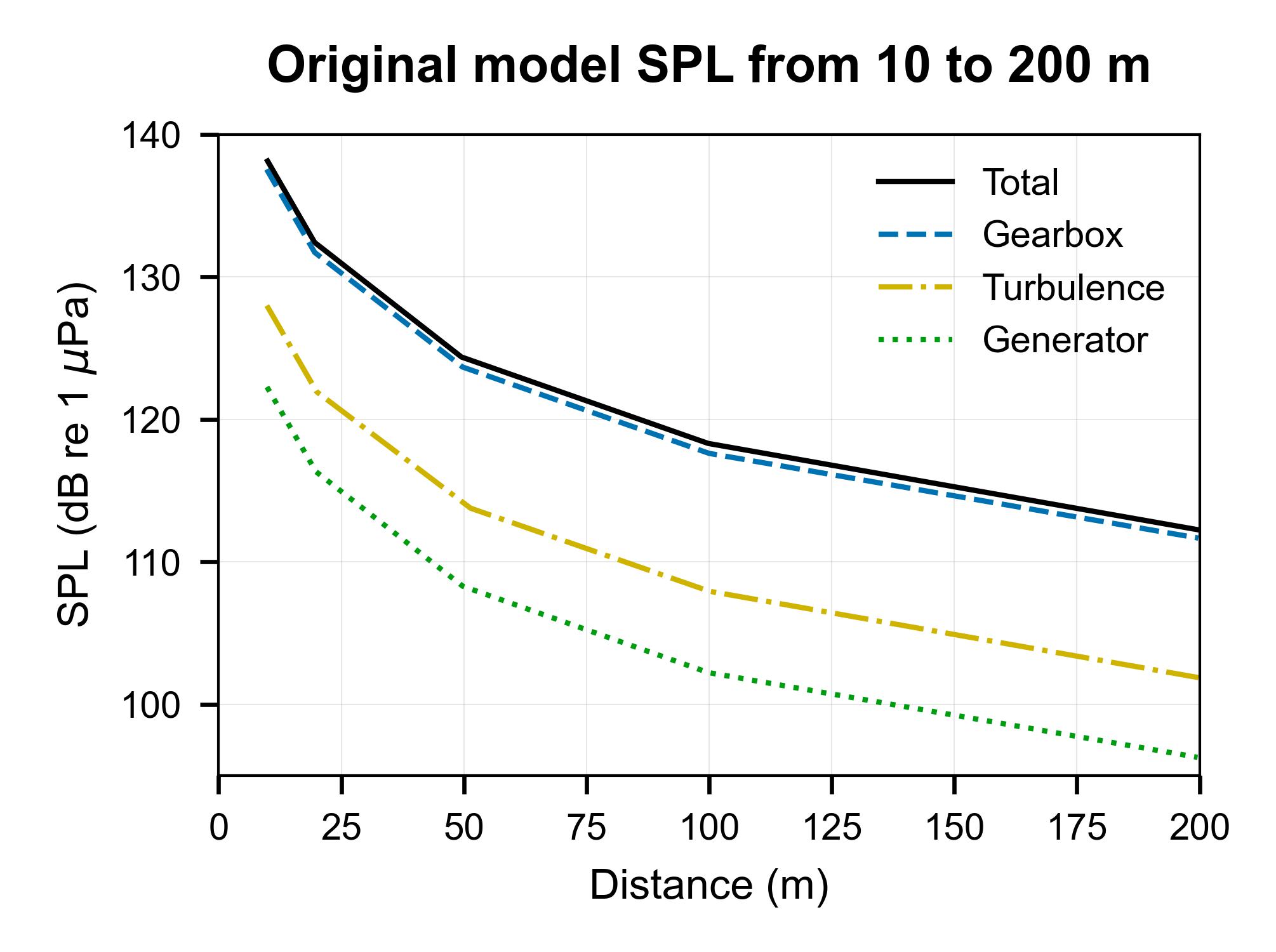}
    \caption{Spatial propagation of acoustic emissions at rated power ($T=7340$ s) from 10 m to 200 m.}
    \label{fig:spl-t-7340}
\end{figure}

Propagation analysis (Fig.~\ref{fig:spl-t-7340}) demonstrates the attenuation of these signals over distance. While the total SPL decays to approximately $112$ dB re 1 $\mu$Pa at 200 m, it remains above the generic auditory threshold for marine mammals within the immediate vicinity of the device ($<100$ m), identifying a critical zone of potential acoustic impact.

\subsection{Influence of Power Electronics Switching Frequency}
The first control intervention involved varying the switching frequency ($F_s$) of the generator-side converter from 1 kHz to 3 kHz. The acoustic simulation results indicated a negligible variation in the aggregate SPL ($<0.05$\% deviation) across the tested range. The high-frequency harmonic noise components generated by the switching events effectively decayed rapidly in the water column and did not contribute significantly to the far-field acoustic pressure.

However, lowering $F_s$ had a substantial detrimental impact on the electrical efficiency of the system. As detailed in Table~\ref{tab:power_electronics_losses}, reducing $F_s$ to 1.5 kHz resulted in a massive increase in harmonic distortion currents, causing thermal losses in the filter components to rise by over 2000\% (from 0.006 kWh to 0.126 kWh over the simulation period). Consequently, manipulating $F_s$ was determined to be an ineffective mitigation strategy, yielding no acoustic benefit while compromising the thermal stability of the power electronics.

\begin{table}[h!]
\caption{Impact of switching frequency ($F_s$) on power electronics energy losses, compared to the reference value (2.5 kHz).}
\label{tab:power_electronics_losses}
\centering
\footnotesize
\setlength{\tabcolsep}{3pt}
\begin{tabularx}{0.8\linewidth}{lXXXX}
\hline
\textbf{Component} & \textbf{2.5 kHz} & \textbf{1.0 kHz} & \textbf{1.5 kHz} & \textbf{3.0 kHz} \\
 & (kWh) & (\% change) & (\% change) & (\% change) \\
\hline
Choke & 0.017 & -15\% & -4\% & +7\% \\
Filter & 0.006 & +988\% & +2107\% & -32\% \\
Transformer & 0.091 & +4\% & +4\% & 0\% \\
\midrule
\textbf{Total Loss} & \textbf{0.114} & \textbf{+50\%} & \textbf{+108\%} & \textbf{0\%} \\
\hline
\end{tabularx}
\end{table}

\subsection{Operational Control via MPPT Tuning}
The second intervention evaluated the de-tuning of the MPPT gain coefficient ($K_{opt}$). Increasing $K_{opt}$ forces the turbine to operate at a lower rotational speed ($\omega$) for a given flow velocity, thereby reducing the tip speed ratio ($\lambda$). Since hydrodynamic noise scales with $U_{tip}^5$ and mechanical noise scales with shaft speed, this strategy effectively reduces source generation.

\begin{figure}[!htbp]
    \centering
    \includegraphics[width=0.6\linewidth]{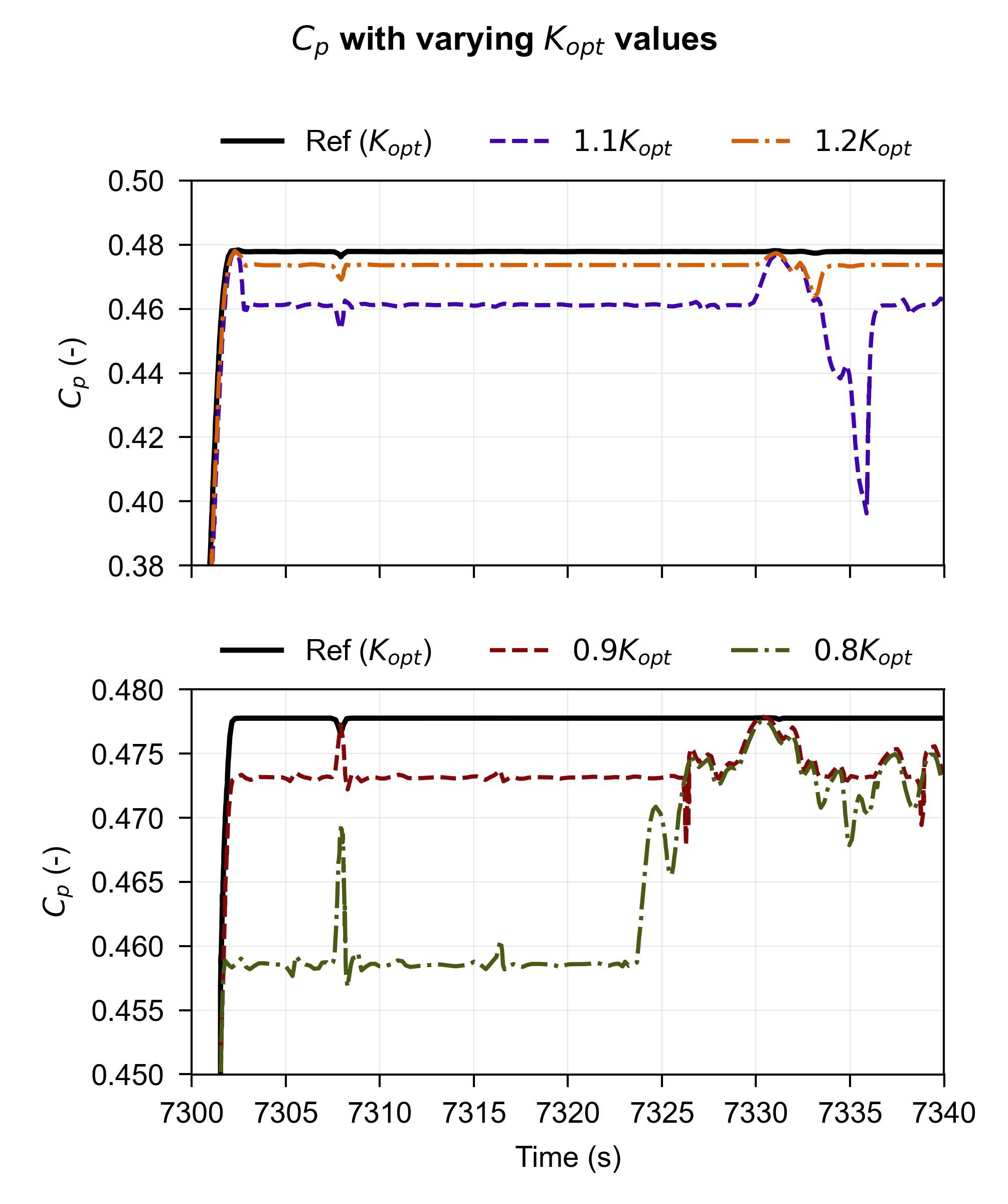}
    \caption{Dynamic step response of the power coefficient ($C_p$) under varying $K_{opt}$ values. The instability in lower $K_{opt}$ regimes highlights the trade-off between acoustic mitigation and control stability.}
    \label{fig:step_response}
\end{figure}

Fig.~\ref{fig:step_response} demonstrates the dynamic stability of the system under different gain coefficients. While lower $K_{opt}$ values (0.8, 0.9) introduced instability in the power coefficient ($C_p$) tracking, the higher values (1.1, 1.2) maintained stable operation while effectively capping the rotational speed.

\begin{figure}[!htbp]
    \centering
    \includegraphics[width=0.75\linewidth]{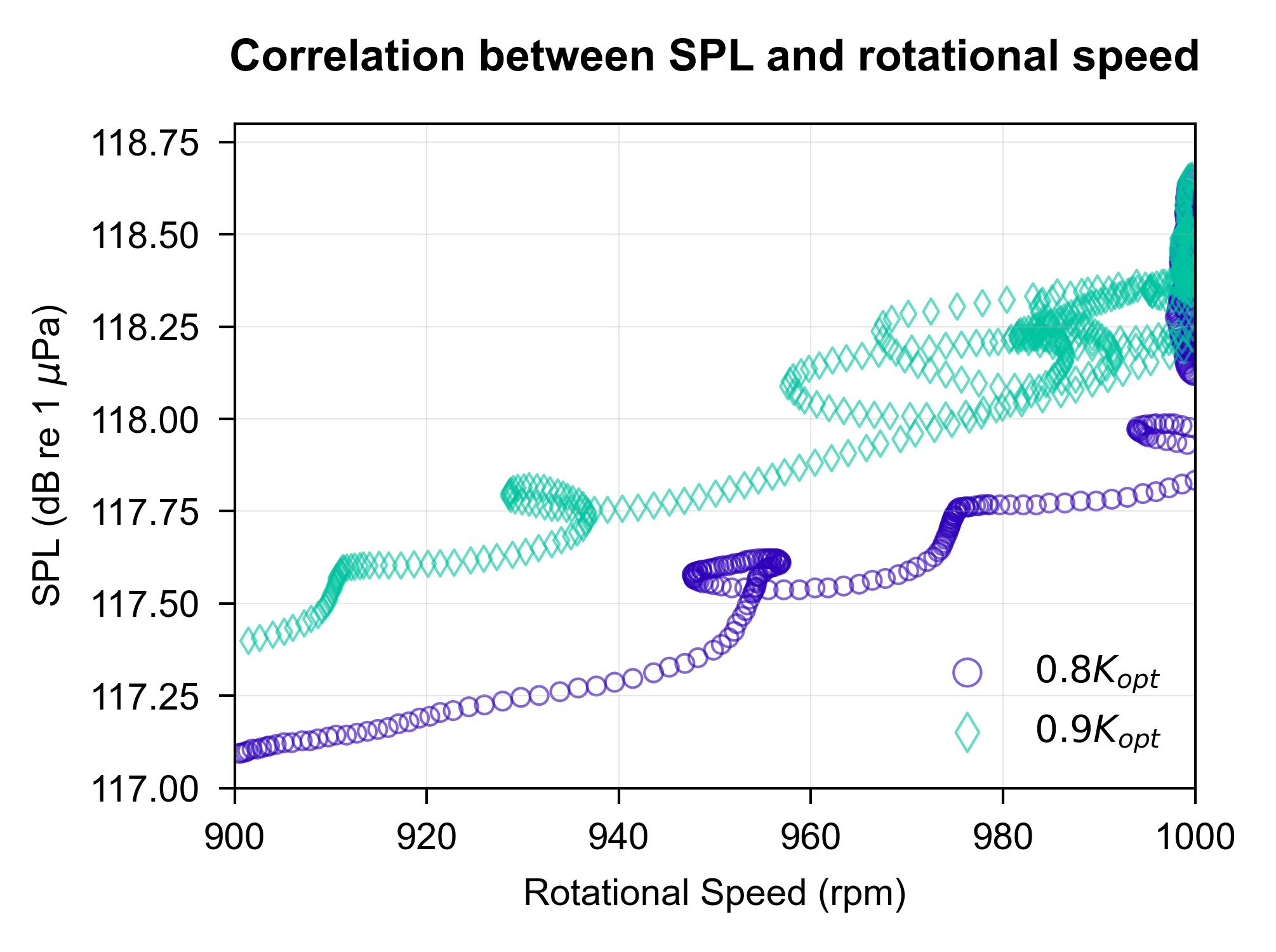}
    \caption{Correlation between SPL and rotational speed under different MPPT gain coefficients ($K_{opt}$). The de-tuned controller ($1.2 K_{opt}$) limits maximum RPM, capping peak acoustic emissions.}
    \label{fig:spl-vs-rpm}
\end{figure}

Fig.~\ref{fig:spl-vs-rpm} illustrates the non-linear relationship between SPL and rotational speed under these regimes. The application of a $1.2 K_{opt}$ factor constrained the rotational speed, keeping the maximum SPL below critical thresholds more frequently than the reference case.

Table~\ref{tab:mppt_tradeoff} quantifies the engineering trade-off associated with this strategy. While increasing $K_{opt}$ to 1.2 resulted in a mean SPL reduction of 0.28\% (on a logarithmic scale, representing a tangible decrease in acoustic energy), it incurred an energy yield penalty of 3.58\%. Conversely, decreasing $K_{opt}$ (0.8) increased both energy production and noise, confirming the direct coupling between aerodynamic efficiency and acoustic emission.

\begin{table}[h!]
\caption{Trade-off between energy yield and acoustic emission under MPPT de-tuning.}
\label{tab:mppt_tradeoff}
\centering
\begin{tabularx}{0.8\linewidth}{lXXX}
\hline
\textbf{Control} & \textbf{Energy Yield} & \textbf{Yield} & \textbf{Mean SPL} \\
\textbf{Factor} & (kWh) & \textbf{Loss (\%)} & \textbf{Variation (\%)} \\
\hline
$K_{opt}$ (Ref) & 8.287 & - & - \\
$0.8 K_{opt}$ & 8.132 & 1.87\% & +0.16\% \\
$1.2 K_{opt}$ & 7.990 & 3.58\% & -0.28\% \\
\hline
\end{tabularx}
\end{table}

\subsection{Impact of Drivetrain Architecture (Direct-Drive)}
\begin{figure}[!htbp]
    \centering
    \includegraphics[width=0.75\linewidth]{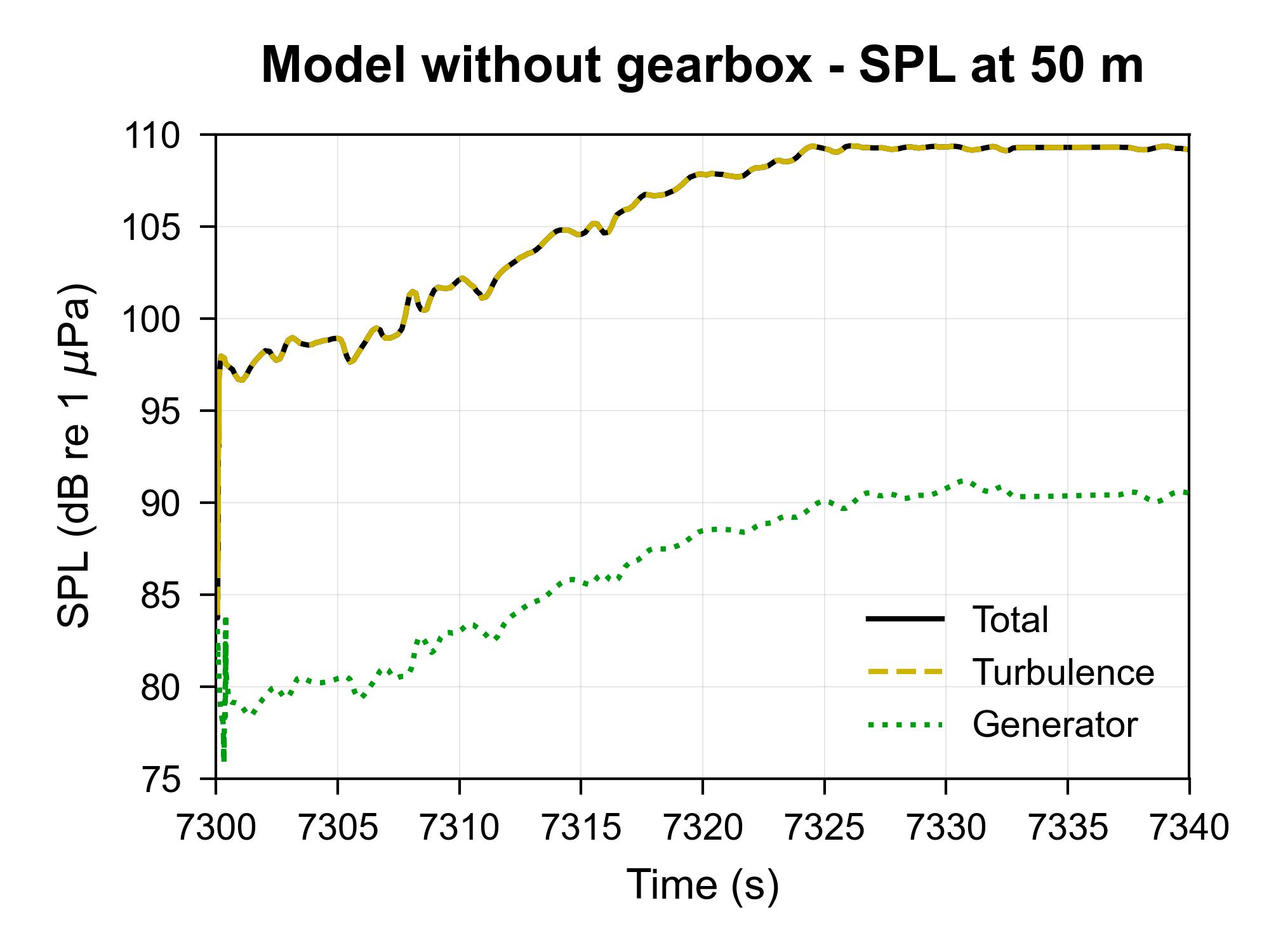}
    \caption{Acoustic profile of the direct-drive PMSG architecture. Total SPL (solid line) is now dominated solely by inflow turbulence (dashed line), significantly lowering the overall noise floor compared to the geared baseline.}
    \label{fig:total-spl-no-gearbox}
\end{figure}

The final evaluation compared the reference geared design with a direct-drive PMSG architecture. As shown in Fig.~\ref{fig:total-spl-no-gearbox}, the removal of the gearbox eliminates the dominant mechanical tonal noise source. The acoustic profile of the direct-drive system is strictly governed by inflow turbulence, resulting in a total SPL reduction of approximately 10 dB re 1 $\mu$Pa at 50 m.

Critically, under the direct-drive topology, the generator noise component (dotted line) remains well below the turbulence noise floor (yellow dashed line), rendering the electromagnetic noise acoustically insignificant. At the rated operation ($T=7340$ s), the direct-drive turbine generated a total SPL of $109.2$ dB re 1 $\mu$Pa, compared to $124.2$ dB re 1 $\mu$Pa for the geared baseline.

\section{Discussion}

\subsection{Acoustic Risk as a Licensing Constraint}
The baseline simulation results quantify the extent of the environmental compliance challenge facing tidal stream developers. With a total source level reaching 124.2 dB re 1 $\mu$Pa, the geared turbine creates a zone of influence where marine mammals are at risk of TTS within a 100 m radius and PTS within 10 m.

From a licensing perspective, this acoustic footprint is significant. In high-energy tidal channels, which are often narrow and serve as migratory bottlenecks for species such as harbour porpoises, a 100 m exclusion zone per turbine, a value consistent with empirical measurements from the SeaGen commercial demonstrator~\cite{turbines2011seagen}, could cumulatively result in an acoustic barrier across the entire channel when scaled to a commercial array.

Furthermore, analysis of the flow speed cumulative distribution function indicates that the turbine operates above the critical acoustic threshold for approximately 56\% of the tidal cycle (Table~\ref{tab:risk_probability}). This high probability of exceedance implies that without active control, the turbine would be non-compliant for the majority of its operational life. Consequently, passive mitigation is insufficient; active engineering intervention is required to secure consenting.

\begin{table}[h!]
\caption{Operational conditions corresponding to the onset of TTS risk ($SPL \ge 116.5$ dB re 1 $\mu$Pa) for marine mammals.}
\label{tab:risk_probability}
\centering
\footnotesize
\begin{tabularx}{0.8\linewidth}{lXXX}
\hline
\textbf{Control} & \textbf{Flow} & \textbf{Risk} & \textbf{Power} \\
\textbf{Strategy} & \textbf{Speed (m/s)} & \textbf{Probability (\%)} & \textbf{Limit (MW)} \\
\hline
Reference & 1.93 & 56.0\% & 0.60 \\
$0.8 K_{opt}$ & 1.87 & 56.5\% & 0.52 \\
$1.2 K_{opt}$ & 2.01 & 53.7\% & 0.67 \\
\hline
\end{tabularx}
\end{table}

\subsection{Evaluation of Operational Mitigation Strategies}
The investigation into control-based mitigation reveals distinct hierarchies of effectiveness.

\subsubsection{Inefficacy of Switching Frequency Control}
Modulating the power electronics switching frequency ($F_s$) proved to be an ineffective strategy. While higher frequencies theoretically push harmonic noise outside the sensitive hearing bandwidth of larger mammals, the simulated reduction in aggregate SPL was negligible ($<1\%$). Conversely, the engineering penalty was severe; lowering $F_s$ to 1.5 kHz increased filter energy losses by over 2000\%. This creates an unacceptable thermal management risk for the power electronics pod, which is often sealed and passively cooled in subsea environments. Therefore, $F_s$ tuning should be reserved for electrical power quality management rather than acoustic mitigation.

\subsubsection{The Cost of Silence: MPPT De-tuning}
In contrast, MPPT coefficient tuning emerged as a viable Tier 2 operational strategy. By increasing the $K_{opt}$ gain to 1.2, the control system effectively caps the rotor's rotational speed during peak tidal flows. This creates a direct trade-off: a 0.28\% reduction in mean noise levels for a 3.58\% loss in annual energy yield.

While a 3.58\% revenue loss is non-trivial, it must be weighed against the alternative of forced shutdowns. If a site license mandates complete curtailment during migration seasons, the generation loss would be 100\% for that period. By utilising MPPT de-tuning, operators can implement a low-noise mode — maintaining 96.4\% of generation capacity while staying compliant with acoustic thresholds. This operational flexibility could be the deciding factor in making a project bankable in environmentally sensitive waters.

\subsection{Design Implications: The Case for Direct-Drive}
The most profound mitigation is achieved at the design stage (Tier 1). The transition to a direct-drive PMSG architecture eliminated the tonal mechanical noise entirely, reducing the source level by 10 dB. By removing the gearbox, the acoustic profile becomes dominated solely by inflow turbulence, which is broadband and less likely to trigger sharp auditory injury responses compared to tonal machinery noise.

Although direct-drive generators are typically heavier and require more expensive rare-earth magnets (increasing CAPEX), this study suggests they offer superior environmental performability. For sites with strict acoustic constraints, the direct-drive topology effectively de-risks the permitting process.

\subsection{A Tiered Mitigation Framework}
Based on these findings, we propose a tiered engineering framework for acoustic compliance:
\begin{itemize}
    \item \textbf{Tier 1 (Design Phase):} For sites identified as critical habitats, direct-drive PMSG architectures should be prioritised to minimise the baseline source level.
    \item \textbf{Tier 2 (Operational Phase):} For geared fleets or transient biological events, MPPT de-tuning should be implemented as a dynamic control state, triggered by real-time passive acoustic monitoring of marine mammal presence.
\end{itemize}

\section{Conclusions}
This study addresses the critical barrier of environmental licensing for tidal current converters (TCCs) by proposing an acoustic-aware control framework. By shifting the focus from passive impact assessment to active engineering intervention, we demonstrated that acoustic emissions can be effectively managed through drivetrain selection and control system tuning.

The simulation results support the following engineering conclusions:
\begin{itemize}
    \item \textbf{Architecture is the Primary Mitigation:} The transition from a geared induction generator to a direct-drive PMSG is the most effective strategy for acoustically sensitive sites. Eliminating the gearbox removed the dominant tonal noise components, resulting in a source level reduction of approximately 10 dB re 1 $\mu$Pa. This design choice effectively de-risks the permitting process for high-value marine habitats.
    \item \textbf{Operational Flexibility via MPPT:} For operational mitigation, de-tuning the MPPT coefficient ($K_{opt}$) offers a viable low-noise mode. Increasing $K_{opt}$ by a factor of 1.2 successfully capped rotational speeds during peak flows, reducing the probability of exceeding marine mammal injury thresholds. This comes at a quantified cost: a 3.58\% reduction in annual energy yield. This trade-off is economically superior to the alternative of total site shutdown during migration periods.
    \item \textbf{Ineffectiveness of $F_s$ Control:} Varying the switching frequency of the power electronics is not a recommended mitigation strategy. The acoustic reduction was negligible ($<1\%$), while the engineering penalty was severe, with filter energy losses increasing by over 2000\% at lower frequencies, posing significant thermal risks to subsea equipment.
\end{itemize}

Ultimately, this work suggests a tiered approach for the tidal industry: prioritise direct-drive topologies in the design phase for noise-constrained environments, and utilise dynamic $K_{opt}$ control as a flexible tool to ensure real-time regulatory compliance. By integrating these strategies, developers can maximise energy extraction while adhering to rigorous environmental standards, thereby accelerating the commercial deployment of marine renewable energy.

\subsection{Future Work}
While this study establishes a foundational framework for acoustic-aware control, several avenues remain for extending this research toward commercial viability and holistic environmental integration.

\begin{itemize}
    \item \textbf{Techno-Economic Analysis of Strategic Derating:}
    While the engineering trade-off of MPPT de-tuning was quantified at a 3.58\% energy yield loss, the long-term financial implications remain to be modelled. Future research should perform a levelised cost of energy sensitivity analysis comparing two scenarios: (A) a high-CAPEX direct-drive fleet operating at full capacity, versus (B) a lower-CAPEX geared fleet utilising seasonal strategic derating. Such case studies would determine the financial break-even point for acoustic compliance in varying regulatory environments.

    \item \textbf{Array-Scale Acoustic Modelling:}
    Commercial projects will involve arrays of multiple turbines. As noted in the foundational thesis work, the acoustic signature of a single device does not linearly scale to an array. Future simulations must account for constructive and destructive interference patterns between turbines, as well as the impact of upstream wake turbulence on the acoustic emissions of downstream rotors. This is critical for assessing the cumulative acoustic barrier effect in narrow tidal channels.

    \item \textbf{Closed-Loop Feedback Control:}
    The current control strategy relies on static parameter tuning ($K_{opt}$). A logical progression is the development of a closed-loop smart mitigation system, where real-time data from passive acoustic monitoring hydrophones serves as an input to the supervisory controller. This would allow the turbine to trigger MPPT curtailment only when marine mammals are detected in the vicinity, minimising energy loss compared to blanket seasonal de-rating.

    \item \textbf{Comprehensive Source Integration:}
    This study focused on hydrodynamic and drivetrain noise. However, secondary sources such as structural vibration resonance, mooring line snapping events, and pile-driving noise during installation also contribute to the total soundscape. Integrating these sources into the Simulink model would provide a complete spectral representation of the TCC lifecycle emissions.
\end{itemize}

\addcontentsline{toc}{section}{Acknowledgements}
\noindent\textbf{\large{Acknowledgements}}
\vspace{0.5cm}

\noindent The authors want to thank the supervisory team for providing valuable guidance and explanations of the critical points within this study.

\cleardoublepage
\phantomsection
\addcontentsline{toc}{section}{References}
\bibliographystyle{IEEEtran} 
\bibliography{MANUSCRIPT} 

\end{document}